\documentclass[11pt,a4paper]{article}
\pdfoutput=1

\usepackage{jheppub}
\usepackage{psfrag}
\usepackage{slashed}
\usepackage{cancel}
\usepackage{lscape}

\usepackage{caption}
\usepackage{array}
\usepackage{graphicx}
\usepackage{subcaption}

\usepackage[utf8]{inputenc}

\definecolor{mygreen}{rgb}{0,0.7,0}

\definecolor{myblue}{rgb}{0,0,0.7}

\definecolor{myred}{rgb}{0.7,0,0}

\def\cN{\mathcal{N}}
\def\cO{\mathcal{O}}
\def\cD{\mathcal{D}}
\def\cQ{\mathcal{Q}}
\def\nn{\nonumber \\ }

\DeclareMathOperator{\tr}{\rm tr}
\def\trm{\tr_-}
\def\trp{\tr_+}

\def\trfive{\tr_5}

\def\braket#1{\langle #1 \rangle}

\def\eps{\epsilon}

\def\usegraph#1#2{\includegraphics[scale=1.0,trim=0 #1 0 0]{graphs/#2.pdf}}

\preprint{Edinburgh 2016/09}

\title{Local integrands for two-loop all-plus Yang-Mills amplitudes}

\author[a]{Simon Badger,}
\author[a]{Gustav Mogull,}
\author[a]{Tiziano Peraro}
\affiliation[a]{
Higgs Centre for Theoretical Physics, School of Physics and Astronomy,\\ The University of
Edinburgh, Edinburgh EH9 3JZ, Scotland, UK%
}
\emailAdd{simon.badger@ed.ac.uk}
\emailAdd{g.mogull@ed.ac.uk}
\emailAdd{tiziano.peraro@ed.ac.uk}

\abstract{We express the planar five- and six-gluon two-loop Yang-Mills amplitudes
with all positive helicities in compact analytic form using $D$-dimensional
local integrands that are free of spurious singularities.
The integrand is fixed from on-shell tree amplitudes in six dimensions using $D$-dimensional generalised unitarity cuts.
The resulting expressions are shown to have manifest infrared behaviour at the integrand level.
We also find simple representations of the rational terms obtained after integration in $4-2\eps$ dimensions.}

\begin{document}
\maketitle
\flushbottom

\section{Introduction}\label{sec:intro}

The continuing development of techniques for calculating scattering amplitudes
is of vital importance for making precision predictions at collider experiments.
The experimental data now being collected at Run II of the LHC will allow the study
of many observables with percent-level uncertainties. This level of precision represents
a serious challenge for current perturbative techniques where a minimum of next-to-next-to-leading
order (NNLO) precision is desirable. This is particularly true for higher multiplicity
final states where the current bottleneck lies in the unknown two-loop matrix elements.

Traditional Feynman diagram techniques have been sufficient for the majority of $2\to2$
matrix elements, yet combining both real and virtual corrections into fully differential NNLO predictions
is by no means a simple task. In the last few years considerable efforts have been made to
develop infrared subtraction methods which have resulted in the majority of $2\to2$ processes
becoming available at NNLO accuracy in QCD. We refer the reader to the recent Les Houches working group report
and references therein for a review of known predictions~\cite{Badger:2016bpw}.

High multiplicity predictions are still restricted to NLO accuracy where algebraic algorithms
have been developed to overcome the increased complexity of the kinematic algebra.
Integrand reduction~\cite{Ossola:2006us}, unitarity~\cite{Bern:1994zx,Bern:1994cg} and generalised
unitarity~\cite{Britto:2004nc,Giele:2008ve} together with both on-shell and off-shell recursive
techniques have been important steps in the construction of automated one-loop amplitude programs.
These techniques have been the starting point for attempts to obtain a fully algebraic approach to
two-loop amplitude computations in non-supersymmetric gauge theories
\cite{Mastrolia:2011pr,Badger:2012dp,Zhang:2012ce,Mastrolia:2012an,Badger:2012dv,Kleiss:2012yv,Feng:2012bm,Mastrolia:2012wf,Mastrolia:2013kca,Badger:2013gxa,Mastrolia:2016dhn,Kosower:2011ty,Larsen:2012sx,CaronHuot:2012ab,Johansson:2012zv,Johansson:2013sda,Johansson:2015ava,Sogaard:2013fpa,Sogaard:2013yga,Sogaard:2014jla,Sogaard:2014oka,Sogaard:2014ila,Ita:2015tya,Larsen:2015ped}.

With a considerable jump in complexity from one to two loops, combined with the difficulty of going beyond $2\to2$
kinematics, it is not clear whether useable expressions (as far as phase space
integration is concerned) would be obtained from using these algorithms as they exist currently.
Two main issues can be identified: there are a larger number of master integrals in comparison to those defined by
integration-by-parts reduction and there are a large number of coefficients containing unphysical singularities.
These issues can result in extremely large analytic expressions and slow numerical evaluation.

Expressing amplitudes in a form manifestly free of these spurious singularities can lead to
remarkably efficient evaluations. This is both because numerical stability is improved but also
because analytic formulae are highly constrained and compact representations can be obtained.
The form of the one-loop five-gluon amplitude obtained by Bern, Dixon and Kosower is an excellent
example of what can be achieved when physical properties and symmetries are manifest
\cite{Bern:1993mq}. The basis of integral functions obtained through the multi-loop integrand reduction procedure
are not unique and often hide the properties of locality and universal infrared behaviour.
If one could select a basis of master integrals with these properties manifest
then the reward could be considerably more compact amplitude expressions.

Progress in calculating amplitudes for supersymmetric Yang-Mills (SYM) theories,
most notably maximally supersymmetric $\cN=4$, has gone in a different direction.
Inspired by Witten's twistor-string theory \cite{Witten:2003nn},
more exotic techniques have been developed that go beyond conventional unitarity-based approaches.
An especially exciting development has been the all-loop integrand for scattering
amplitudes in the planar sector of $\cN=4$ \cite{ArkaniHamed:2010kv,ArkaniHamed:2010gh}.
This generalises the BCFW recursion relations for tree amplitudes \cite{Britto:2004ap,Britto:2005fq} to loop level,
building amplitudes out of chiral integrals with unit leading singularities.
Amongst many other important properties, the presentation is local and makes use of infrared (IR) finite integrals. This technique has been
applied to a variety of explicit cases, most recently to all multiplicity two-loop amplitudes in
planar $\cN=4$~\cite{Bourjaily:2015jna}. One naturally questions whether such an approach could
be applied in pure Yang-Mills theory where it is conventional to work in $D\neq4$.

In this article we investigate the possibility of obtaining compact two-loop QCD amplitudes free of
spurious singularities using integrand reduction and generalised unitarity cuts in $D$-dimensions.
Rather than constructing $D$-dimensional local integrands from first principles,
we focus on their uses in the all-plus sector of Yang-Mills theory.
Since the all-plus Yang-Mills amplitudes are closely related to those in $\cN=4$ SYM,
it is a useful starting point that will allow us to directly recycle the supersymmetric expressions.
The result will be a partially-determined basis of master integrals onto which we can fit cut solutions.
Specifically, we will consider the planar five- and six-gluon all-plus amplitudes at two loops,
integrated expressions for which have now been written down by Gehrmann,
Henn and Lo Presti \cite{Gehrmann:2015bfy} and Dunbar and Perkins
\cite{Dunbar:2016aux,Dunbar:2016cxp,Dunbar:2016gjb}.
The resulting amplitude expressions are free of unphysical poles before integration
and have a remarkably simple infrared structure.

The rest of our paper is organised as follows.
In section \ref{sec:oneloop} we introduce $D$-dimensional local integrands by considering a box integral,
then use the structure to rewrite the planar five- and six-point one-loop all-plus amplitudes.
In section \ref{sec:unitarity} we develop the $D$-dimensional unitarity
techniques that we will use to fit cut solutions.
section \ref{sec:twoloops} contains our main results:
the planar five- and six-point two-loop all-plus amplitudes in local-integrand form.
We study the infrared divergences of these in section \ref{sec:ir}
and then present the integrated six-point rational terms in section \ref{sec:rational}.
Finally, in section \ref{sec:conclusions} we draw our conclusions and discuss future directions.

\subsection{Notation}

Throughout this paper we will work in dimensional regularisation
with different dimensions in play simultaneously:
\begin{description}
\item[$D$] The number of dimensions in dimensional regularisation, $D=4-2\epsilon$.
\item[$D_s$] The spin dimension of internal gluons.
\item[$\cD$] The number of dimensions in which we embed $D$-dimensional momenta, $\cD=6$.
\end{description}
One can obtain results in the 't Hooft Veltman (tHV) scheme by setting $D_s=4-2\eps$ and the the four dimensional
helicity scheme by setting $D_s=4$ \cite{Bern:2002zk}.

We will adopt conventional notation for spinor and Lorentz products.
Four-dimensional external momenta will be denoted $p_i$ and we will use the usual shorthands
\begin{align}
p_{ij\cdots k}=p_i+p_j+\cdots+p_k,&&
s_{ij\cdots k}=p_{ij\cdots k}^2.
\end{align}
Spinor products will be constructed from holomorphic ($\lambda_a$) and anti-holomorphic
($\tilde{\lambda}_{\dot{a}}$) two-component Weyl spinors,
such that $\braket{ij}=\lambda_{i,a}\lambda_j^a$
and $[ij]=\tilde{\lambda}_i^{\dot{a}}\tilde{\lambda}_{j,\dot{a}}$
We will also find it convenient to write local integrands in terms of traces over $\gamma$-matrices,
\begin{align}
\text{tr}_\pm(ij\cdots k)
=\frac{1}{2}\tr((1\pm\gamma_5)\slashed{p}_i\slashed{p}_j\cdots\slashed{p}_k),&&
\trfive(ij\cdots k)=\tr(\gamma_5\slashed{p}_i\slashed{p}_j\cdots\slashed{p}_k),
\end{align}
where $\trfive(ijkl)=4i\epsilon_{\mu\nu\rho\sigma}p_i^\mu p_j^\nu p_k^\rho p_l^\sigma$ and,
for instance, $\trp(ijkl)=[ij]\braket{jk}[kl]\braket{li}$.

$D$-dimensional loop momenta $\ell_i$ will sometimes be separated into their four-dimensional
and $(-2\epsilon)$-dimensional parts $\ell_i=\bar{\ell_i}+\ell_i^{[-2\epsilon]}$.
Rotational invariance in the extra dimensions forces $\ell_i^{[-2\epsilon]}$
to appear in the combinations $\mu_{ij}=-\ell_i^{[-2\epsilon]}\cdot\ell_j^{[-2\epsilon]}$.
For one-loop integrals we will write $\mu^2=-(\ell^{[-2\epsilon]})^2$.
These loop momenta will often be included in Dirac traces,
necessitating the use of a $D$-dimensional Clifford algebra.
Formally our approach will follow that used by 't Hooft and Veltman \cite{Collins:1984xc,'tHooft:1972fi}
and elaborated in in ref.~\cite{Bern:1995db} but in practice
we will overcome such ambiguities by decomposing into four-dimensional traces
\begin{align}
\text{tr}_\pm(i_1\cdots i_k\ell_x\ell_yi_{k+1}\cdots i_n)
=\text{tr}_\pm(i_1\cdots i_k\bar{\ell}_x\bar{\ell}_yi_{k+1}\cdots i_n)
-\mu_{xy}\text{tr}_\pm(i_1\cdots i_n).
\end{align}
This decomposition will allow us to evaluate the $D$-dimensional traces using
\begin{align}\label{eq:trstacking}
\text{tr}_\pm(i_1\cdots i_k\ell_x\ell_yi_{k+1}\cdots i_n)
=\frac{\text{tr}_\pm(i_1\cdots i_n)}{s_{i_k,i_{k+1}}}\text{tr}_\pm(i_k\ell_x\ell_yi_{k+1}),
\end{align}
where $k$ should in this case be odd.

Finally, our integral conventions will be as follows.
For a given diagram topology $T$ defined by a set of $m$ massless propagators $\{\cQ_\alpha\}$
the one-loop integration operator will be
\begin{subequations}
\begin{align}\label{eq:1loopintegral}
I^D_T\left[\mathcal{P}(p_i,\ell,\mu^2)\right]
\equiv i(-1)^{m+1}(4\pi)^{D/2} \mu_R^{4-D} \int\!\frac{d^D\ell}{(2\pi)^D}
\frac{\mathcal{P}(p_i,\ell,\mu^2)}
{\prod_\alpha \cQ_{\alpha}(p_i,\ell)},
\end{align}
and the two-loop integration operator will be
\begin{align}\label{eq:2loopintegral}
I^D_T\left[\mathcal{P}(p_i,\ell_i,\mu_{ij})\right]
\equiv-(4\pi)^D \mu_R^{2(4-D)} \int\!\frac{d^D\ell_1d^D\ell_2}{(2\pi)^{2D}}
\frac{\mathcal{P}(p_i,\ell_i,\mu_{ij})}
{\prod_\alpha \cQ_{\alpha}(p_i,\ell_1,\ell_2)},
\end{align}
\end{subequations}
where $\mu_R^2$ is the regularisation scale and throughout the rest of the article we will set
$\mu_R^2=1$. In general we will identify topologies $T$ by explicitly drawing them.\footnote{ Notice that the conventions differ between one- and two-loop integrals.}

\section{One-loop local integrands}\label{sec:oneloop}

\subsection{The box integral in $D=4-2\epsilon$}

To motivate our discussion of local integrands we begin by considering the scalar box integral
\begin{align}\label{eq:boxintegral}
I^D\bigg(\usegraph{10}{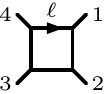}\bigg)
=\frac{r_\Gamma}{st}\left(\frac{2}{\epsilon^2}\left((-s)^{-\epsilon}+(-t)^{-\epsilon}\right)
-\ln^2\left(\frac{s}{t}\right)-\pi^2\right)+\cO(\epsilon),
\end{align}
where $s=s_{12}$ and $t=s_{23}$ are Mandelstam invariants and we have introduced
the standard loop prefactor $r_\Gamma=\Gamma(1+\epsilon)\Gamma^2(1-\epsilon)/\Gamma(1-2\epsilon)$.
The $\epsilon$ pole structure is entirely due to IR divergences from soft regions,
such as when $\ell\to0$ or $\ell-p_1\to0$, and collinear regions,
such as when $\ell$ approaches collinearity with $p_1$.
The possibility of simultaneous soft and collinear divergences gives leading poles at $\cO(\epsilon^{-2})$.

The box integral can be rendered finite by introducing a local numerator\footnote{
    The ``wavy line'' notation was first introduced by Arkani-Hamed et al. to denote
    local integrands in maximally supersymmetric $\mathcal{N}=4$ \cite{ArkaniHamed:2010kv,ArkaniHamed:2010gh}.
    This connection will be explored in the next section.}
\begin{align}\label{eq:localboxintegral}
I^D\bigg(\usegraph{10}{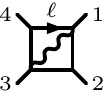}\bigg)\equiv
I^D\bigg(\usegraph{10}{box}\bigg)\left[\trp(1(\ell-p_1)(\ell-p_{12})3)\right].
\end{align}
This numerator vanishes in all of the soft and collinear regions.
To see explicitly that the new integral is finite we evaluate the Dirac trace
\begin{align}
&\trp(1(\ell-p_1)(\ell-p_{12})3)
=\frac{1}{2}\tr(1(\ell-p_1)(\ell-p_{12})3)+\frac{1}{2}\trfive(1(\ell-p_1)(\ell-p_{12})3)
\nonumber\\&\qquad
=\frac{st}{2}-\frac{t}{2}\ell^2-\frac{s}{2}(\ell-p_1)^2-\frac{t}{2}(\ell-p_{12})^2-\frac{s}{2}(\ell+p_4)^2
-\frac{1}{2}\trfive(123\ell).
\end{align}
The spurious $\trfive$ term integrates to zero.
When the propagators are cancelled against their counterparts in the denominator
the new box integral becomes a linear combination of scalar box and triangle integrals\footnote{
  Extra minus signs originate from the one-loop integral convention given in eq.~(\ref{eq:1loopintegral}).}
\begin{align}
I^D\bigg(\usegraph{10}{boxs}\bigg)
&=\frac{st}{2}I^D\bigg(\usegraph{10}{box}\bigg)
\!+\!\frac{s}{2}I^D\bigg(\!\usegraph{9}{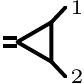}\bigg)\!+\!\frac{t}{2}I^D\bigg(\!\usegraph{9}{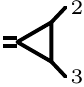}\bigg)
\!+\!\frac{s}{2}I^D\bigg(\!\usegraph{9}{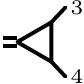}\bigg)\!+\!\frac{t}{2}I^D\bigg(\!\usegraph{9}{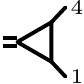}\bigg)
\nonumber\\
&=-\frac{r_\Gamma}{2}\left(\ln^2\left(\frac{s}{t}\right)+\pi^2\right)+\cO(\epsilon).
\end{align}
Here we have used the scalar triangle integral
\begin{align}\label{eq:triintegral}
I^D\bigg(\!\usegraph{9}{tri12}\bigg)=\frac{r_\Gamma}{\epsilon^2}(-s)^{-1-\epsilon}.
\end{align}
The regulated box integral can also be evaluated using a dimension shift \cite{Bern:1992em,Bern:1993kr}
\begin{align}
I^D\bigg(\usegraph{10}{boxs}\bigg)
=(-1+2\epsilon)uI^{D+2}\bigg(\usegraph{10}{box}\bigg),
\end{align}
where $u=s_{13}$.
IR finiteness follows trivially as the box integral is finite in six dimensions.

\subsection{One-loop all-plus amplitudes}\label{sec:oneloopamplitudes}

To continue our motivation of local integrands we now rewrite the five- and six-gluon one-loop all-plus amplitudes in $D=4-2\epsilon$ \cite{Bern:1993qk,Mahlon:1993fe} into such a representation.
We consider only the leading-colour component
\begin{align}\label{eq:1lcolourdecomp}
&\mathcal{A}^{(1)}(1^+,2^+,\cdots,n^+)\nonumber\\
&\qquad=g^nN_c\sum_{\sigma\in S_5}\tr(T^{a_{\sigma(1)}}T^{a_{\sigma(2)}}\cdots T^{a_{\sigma(n)}})
A^{(1)}(\sigma(1^+),\sigma(2^+),\cdots,\sigma(n^+))+\cO(N_c^0),
\end{align}
where $T^a$ are the generators of $\text{SU}(N_c)$ and $g$ is the strong coupling constant.

Our starting point is the local all-loop integrand for scattering amplitudes
in the planar MHV sector of $\cN=4$ SYM \cite{ArkaniHamed:2010kv,ArkaniHamed:2010gh}.
These supersymmetric expressions are by now quite familiar and,
amongst many other interesting properties, they are known to exhibit simple IR behaviour.
To use them in the context of all-plus Yang-Mills theory
we recall that the corresponding one-loop amplitudes are related by a dimension shift \cite{Bern:1996ja}.
Before integration this is equivalent to replacing the supersymmetric delta function
$\delta^8(Q)$ with $(D_s-2)\mu^4$ in the all-plus integrand.

The momentum-twistor formalism used to write the $\cN=4$ expressions seemingly
ties them to four dimensions so we begin by translating to a manifestly $D$-dimensional language.
Full details of the procedure up to two loops are given in appendix \ref{sec:N4localintegrands};
for now we merely notice that, at one loop,
pentagon integrals always operate on twistor brackets involving loop momenta.
These twistor brackets are, up to a helicity-dependent scaling,
equivalent to Dirac traces with a positive projector:
\begin{align}\label{eq:squiggledef}
\usegraph{13}{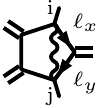}\sim\braket{AB|(i-1,i,i+1)\cap(j-1,j,j+1)}\sim\trp(i\ell_x\ell_yj).
\end{align}
The only other four-dimensional one-loop integrals are scalar box integrals.

With this correspondence in mind we define $D$-dimensional regulated pentagon integrals as
\begin{align}\label{eq:regulatedpentagon}
I^D\bigg(\!\usegraph{13}{pentschemes}\!\bigg)\left[\mathcal{P}(p_i,\ell,\mu^2)\right]
\equiv I^D\bigg(\!\usegraph{13}{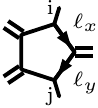}\!\bigg)
\left[\trp(i\ell_x\ell_yj)\mathcal{P}(p_i,\ell,\mu^2)\right],
\end{align}
which is the same definition as we made for the regulated box in eq.~(\ref{eq:localboxintegral}).
This ``wavy line'' notation is not quite the same as that used by Arkani-Hamed et al.
but it has the same property of controlling IR divergences.

The one-loop all-plus amplitudes can now be re-expressed as
\begin{subequations}\label{eq:1lallplus}
\begin{align}
&A^{(1)}(1^+,2^+,3^+,4^+,5^+)=
-\frac{i(D_s-2)}{\braket{12}\braket{23}\braket{34}\braket{45}\braket{51}(4\pi)^{2-\epsilon}}
\times\label{eq:5pt1lallplus}\\
&\qquad
\Bigg(\frac{\trp(1345)}{s_{13}}I^D\bigg(\!\usegraph{12}{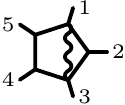}\!\bigg)[\mu^4]
+s_{23}s_{34}I^D\bigg(\usegraph{10}{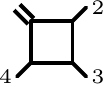}\bigg)[\mu^4]
+s_{12}s_{15}I^D\bigg(\usegraph{10}{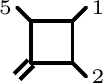}\bigg)[\mu^4]\Bigg),\nonumber\\
&A^{(1)}(1^+,2^+,3^+,4^+,5^+,6^+)=
-\frac{i(D_s-2)}{\braket{12}\braket{23}\braket{34}\braket{45}\braket{56}\braket{61}(4\pi)^{2-\epsilon}}
\times\nonumber\\
&\qquad\Bigg(
\trp(123456)I^D\bigg(\usegraph{12}{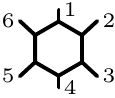}\bigg)[\mu^6]
+\frac{\trp(1456)}{s_{14}}I^D\bigg(\!\usegraph{12}{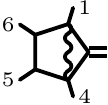}\bigg)[\mu^4]\label{eq:6pt1lallplus}\\
&\qquad\qquad
+\frac{\trp(13(4\!+\!5)6)}{s_{13}}I^D\bigg(\!\usegraph{12}{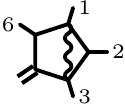}\!\bigg)[\mu^4]
+\frac{\trp(245(6\!+\!1))}{s_{24}}I^D\bigg(\!\usegraph{12}{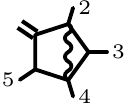}\!\bigg)[\mu^4]\nonumber\\
&\qquad\qquad
+s_{12}s_{61}I^D\bigg(\usegraph{10}{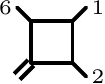}\bigg)[\mu^4]
+s_{23}s_{345}I^D\bigg(\,\usegraph{10}{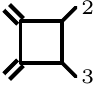}\bigg)[\mu^4]
+s_{34}s_{45}I^D\bigg(\usegraph{10}{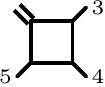}\bigg)[\mu^4]\Bigg).\nonumber
\end{align}
\end{subequations}
The hexagon integral, being a manifestly $D$-dimensional contribution,
is not present in the four-dimensional $\cN=4$ local integrand presentation.
We obtained its value from the $D$-dimensional presentation
of the (parity-even) part of the same $\cN=4$ amplitude given in ref.~\cite{Bern:2008ap}.
We have checked that the above expressions agree with those previously obtained by Bern et
al.~\cite{Bern:1996ja} after integration up to terms of $\mathcal{O}(\epsilon)$.
The main difference here is that, where the previous representation made use of $\mu^6$ pentagons,
we instead use Dirac traces - this keeps power counting of loop momentum
within the expectations of pure Yang-Mills theory.
The relevant one-loop integrals in the $\epsilon\to0$ limit are
\begin{align}
I^D\bigg(\usegraph{10}{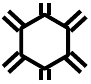}\bigg)[\mu^6]\to0,\qquad
I^D\bigg(\usegraph{13}{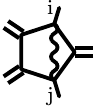}\bigg)[\mu^4]\to\frac{s_{ij}}{6},\qquad
I^D\bigg(\usegraph{10}{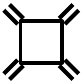}\bigg)[\mu^4]\to-\frac{1}{6}.
\end{align}
Spurious poles in $s_{ij}$ associated with pentagons all cancel.

One can also infer this cancellation of spurious poles, and therefore the locality of the amplitudes,
to all orders in $\epsilon$ by considering the unintegrated expressions.
For instance, in the pentagon integral from eq.~(\ref{eq:5pt1lallplus}) the identity (\ref{eq:trstacking}) gives
\begin{align}
\trp(1(\ell-p_1)(\ell-p_{12})345)=\frac{\trp(1345)}{s_{13}}\trp(1(\ell-p_1)(\ell-p_{12})3).
\end{align}
Similar relationships are applicable to the six-point pentagon integrals.
Therefore the integrands are all local despite there being
unphysical kinematic variables in the denominators.

\section{$D$-dimensional unitarity at two loops}\label{sec:unitarity}

The two-loop results presented in this paper are obtained using $D$-dimensional
integrand reduction and generalised unitarity cuts.  A generic contribution
to a loop amplitude can be expressed, by means of integrand
reduction methods, as a sum of irreducible integrands
  \begin{align}
    \frac{\mathcal{P}(p_i,\ell_i,\mu_i)}
{\prod_{\alpha} \cQ_{\alpha}(p_i, \ell_i)} = \sum_{T} \frac{\Delta_{T}(p_i,\ell_i)}{\prod_{\alpha\in
T} \cQ_{\alpha}(p_i,\ell_i)},
  \end{align}
where the sum $T$ runs over all the subtopologies of the
parent topology.  The irreducible numerators
$\Delta_{T}(p_i,\ell_i)$ can be cast as linear combinations of
loop-momentum-dependent basis elements $m_{T,j}$
\begin{align} \label{eq:deltaparametrization}
  \Delta_{T}(p_i,\ell_i) = \sum_j c_{T,j}(p_j) \, m_{T,j}(p_i,\ell_j),
\end{align}
where values of the unknown coefficients $c_{T,j}$ can be found by
evaluating the integrand on values of the loop momenta satisfying
the multiple-cut conditions $\{\cQ_\alpha = 0, \alpha \in T\}$.
More explicitly,
  \begin{align}
    \Delta_{T}(p_i,\ell_i) =     \left(\frac{\mathcal{P}(p_i,\ell_i,\mu_i)}
{\prod_{\beta \not \in T} \cQ_{\beta}(p_i,\ell_i)} - \sum_{T' \supset T}
\frac{\Delta_{T'}(p_i,\ell_i)}{\prod_{\beta\in T' \setminus T} \cQ_{\beta}(p_i,\ell_i)}\right),
\quad {\textrm {if } \cQ_\alpha = 0 , \alpha \in T}.
  \end{align}

When the propagators are taken on shell, the cut integrand $\mathcal{P}$ factorises into
a product of tree-level amplitudes. These tree-level amplitudes must
be evaluated in $\mathcal{D}>4$ in order to extract the $\mu_{ij}$ terms.
At two loops the minimum embedding dimension is six and we make use of the
six-dimensional spinor-helicity formalism~\cite{Cheung:2009dc,Bern:2010qa,Davies:2011vt}.
The $\cD=6$ dimensional cuts can be dimensionally reduced to
the tHV or four-dimensional helicity (FDH) schemes by considering additional scalar
loops~\cite{Bern:2002zk}. At two loops we include $D_s-\cD$ contributions with a single scalar loop
and $(D_s-\cD)^2$ contributions with two scalar loops,
\begin{align}
  \Delta_T = \Delta_T^{(g,\cD)} + (D_s-\cD)\, \Delta_T^{(s,\cD)} + (D_s-\cD)^2\, \Delta_T^{(s^2,\cD)}.
\end{align}
The full reduction procedure starts from the top-level topology and recursively proceeds to lower
topologies. At each step, the previously computed irreducible numerators are used
to remove poles appearing in the cut numerator.

The external kinematics can be conveniently parameterised in terms of momentum-twistor
variables~\cite{Hodges:2009hk}. In our implementation the six-dimensional spinors are evaluated directly
in terms of the explicit parameterisation given in ref.~\cite{Badger:2016uuq}. In the case of the six-gluon tree-level
amplitudes appearing in this work, we obtained the results through BCFW recursion relations.\footnote{
	We are grateful to Christian Br\o{}nnum-Hansen for providing his Mathematica code for the
	evaluation of 6D amplitudes via BCFW recursion.}
This approach is particularly convenient since it can be applied both numerically or analytically.
We have made use of the ability to evaluate using rational numerics to reconstruct the full analytic form
of the cuts in cases where factorisation of intermediate polynomials became computationally expensive.

The choice for the terms $m_{T,j}(p_i,\ell_j)$ appearing in eq.~\eqref{eq:deltaparametrization} is
not unique and the only requirement they must satisfy is to be independent of each other modulo
the cut propagators $\cQ_\alpha$. In the standard approach a basis of such elements is obtained
via multi-variate polynomial division. While this method is relatively straightforward to
implement, the resulting representation for each irreducible numerator will have
unphysical poles and infrared factorisation properties.

In this paper we use the local integrand structures introduced in the previous section
to construct a set of basis elements $m_{T,j}(p_i,\ell_j)$ which make some properties of the
amplitudes manifest.  In addition to ensuring off-shell symmetries we follow a set of guidelines:
\begin{itemize}
\item the infrared pole structure of the amplitude should follow from its integrand
  representation,
\item the integrand should not contain spurious singularities with
  respect to the external invariants,
\item an $n$-point integrand should match the $(n-1)$-point result
  when taking soft limits of the external particles.
\end{itemize}
We will show how the infrared poles can be extracted from the integrand in section
\ref{sec:softlimits}. As we shall see, this leads to an integrand form for the all-plus
amplitudes presented here with a significantly lower number of
non-vanishing terms compared to the one obtained with a more
traditional polynomial-division-based approach.

\section{Two-loop local integrands}\label{sec:twoloops}

In this section we present the leading-colour parts of the two-loop
all-plus five- and six-gluon scattering amplitudes:
\begin{align}\label{eq:2lcolourdecomp}
&\mathcal{A}^{(2)}(1^+,2^+,\cdots,n^+)\nonumber\\
&=g^{n+2}N_c^2\sum_{\sigma\in S_5}\tr(T^{a_{\sigma(1)}}T^{a_{\sigma(2)}}\cdots T^{a_{\sigma(n)}})
A^{(2)}(\sigma(1^+),\sigma(2^+),\cdots,\sigma(n^+))+\cO(N_c),
\end{align}
where once again $T^a$ are the generators of $\text{SU}(N_c)$ and $g$ is the strong coupling constant.
The five-gluon amplitude was already written down in ref.~\cite{Badger:2013gxa};
here we show how a local integrand presentation allows us to write this result
more compactly and eliminate unphysical poles before integration. The master integrals for
planar $2\to3$ scattering are also now available in analytic
form~\cite{Gehrmann:2015bfy,Papadopoulos:2015jft}.

The two amplitudes are given as sums of integrals of irreducible numerators $\Delta_T$:
\begin{align}
A^{(2)}(1^+,2^+,\cdots,n^+)
=-\frac{i}{\braket{12}\braket{23}\cdots\braket{n1}(4\pi)^{4-2\epsilon}}\sum_TI^D_T\left[\Delta_T\right],
\end{align}
where the two-loop integration operator $I_T^D$ was defined in eq.~(\ref{eq:2loopintegral}).
The sum on $T$ runs over a complete set of two-loop topologies,
many being duplicates of the same diagrams summed over different cyclic orderings.
All genuine two-loop topologies carry the $D$-dimensional prefactor
\begin{align}\label{eq:F1def}
&F_1(\ell_1^{[-2\epsilon]},\ell_2^{[-2\epsilon]})\nonumber\\
&\qquad
=(D_s-2)(\mu_{11}\mu_{22}+(\mu_{11}+\mu_{22})^2
+2\mu_{12}(\mu_{11}+\mu_{22}))+16(\mu_{12}^2-\mu_{11}\mu_{22}),
\end{align}
while one-loop-squared graphs are split into terms proportional to $D_s-2$ and $(D_s-2)^2$,
\begin{subequations}
\begin{align}
F_2(\ell_1^{[-2\epsilon]},\ell_2^{[-2\epsilon]})
&=4(D_s-2)\mu_{12}(\mu_{11}+\mu_{22}),\label{eq:F2def}\\
F_3(\ell_1^{[-2\epsilon]},\ell_2^{[-2\epsilon]})
&=(D_s-2)^2\mu_{11}\mu_{22}.\label{eq:F3def}
\end{align}
\end{subequations}
In all cases the spurious $F_2$ terms integrate to zero.

\subsection{The five-gluon integrand}

The five-point numerators are
\begin{subequations}\label{eq:2L5g}
\begin{align}
\Delta\bigg(\usegraph{9}{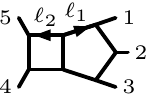}\bigg)
&=s_{45}\trp(1(\ell_1-p_1)(\ell_1-p_{12})345)F_1,
\label{eq:431}\\
\Delta\bigg(\usegraph{13}{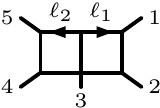}\bigg)
&= -s_{12}s_{45}s_{15}F_1,\label{eq:3315L}\\
\Delta\bigg(\usegraph{13}{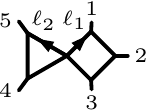}\!\bigg)
&=\trp(1(\ell_1-p_1)(\ell_1-p_{12})345)
\left(F_2+F_3\frac{s_{45}+(\ell_1+\ell_2)^2}{s_{45}}\right),
\label{eq:430}\\
\Delta\bigg(\usegraph{13}{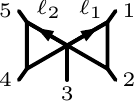}\bigg)
&=\trp(1245)(F_2+F_3)+\frac{F_3}{s_{12}s_{45}}
\bigg(\trp(123\ell_1\ell_2345)
\nonumber\\&\qquad
+(s_{12}s_{45}s_{15}+(s_{12}+s_{45})\trp(1245))(\ell_1+\ell_2)^2\bigg),
\label{eq:3305L}
\end{align}
\end{subequations}
of which the first two are genuine two-loop topologies.
Two numerators from the original representation of ref.~\cite{Badger:2013gxa} are now zero:
these are the one-mass double box and one-mass double triangle.
Their contributions have been absorbed into other topologies.

In this new presentation, both genuine two-loop topologies match their counterparts from the $\cN=4$ all-loop integrand.
The procedure for translating these supersymmetric expressions to the $D$-dimensional language used above is the same as that outlined
in section \ref{sec:oneloopamplitudes} and elaborated in appendix \ref{sec:N4localintegrands};
we see that the supersymmetric delta function $\delta^8(Q)$ is now replaced with the $D$-dimensional prefactor $F_1$.\footnote{ The matching of numerators in $\cN=4$ SYM to all-plus Yang Mills has been conjectured at one loop in ref.~\cite{Bern:1996ja} and at two loops in ref.~\cite{Badger:2013gxa}.  In this paper we have used this property to construct ansätze which are then explicitly checked on the multiple cuts.}
However, we also notice that, even when there are no supersymmetric counterparts to the diagrams,
the same local integrands involving Dirac $\trp$ objects continue 
to be a useful means of expressing loop-momentum dependence.

When forming integrated expressions the rearrangement of Dirac $\trp$ objects demonstrated in
eq.~(\ref{eq:trstacking}) allows the use regulated pentabox and box-triangle integrals:
\begin{subequations}
\begin{align}
I^D\bigg(\usegraph{9}{431}\bigg)\left[\Delta\bigg(\usegraph{9}{431}\bigg)\right]
&=\frac{s_{45}\trp(1345)}{s_{13}}I^D\bigg(\usegraph{9}{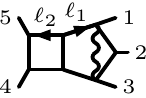}\bigg)[F_1],\\
I^D\bigg(\usegraph{13}{430}\!\bigg)\left[\Delta\bigg(\usegraph{13}{430}\!\bigg)\right]
&=\frac{\trp(1345)}{s_{13}}I^D\bigg(\usegraph{13}{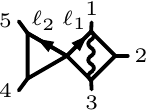}\!\bigg)[F_3]\nonumber\\
&\qquad+\frac{\trp(1345)}{s_{13}s_{45}}I^D\bigg(\usegraph{13}{430s}\!\bigg)[F_3(\ell_1+\ell_2)^2].
\end{align}
\end{subequations}
The two-loop ``wavy line'' notation used here follows precisely the same one-loop conventions
introduced in eqs.~(\ref{eq:localboxintegral}) and (\ref{eq:regulatedpentagon}).

As a final remark on this local representation we notice that the soft limits of the
irreducible numerators match directly onto the four-point numerators.
For example,
\begin{subequations}\label{eq:5ptsoftlimits}
\begin{align}
\Delta\bigg(\usegraph{13}{430}\!\bigg)
&\xrightarrow{p_2\to 0}\Delta\bigg(\usegraph{10}{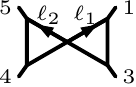}\!\bigg),\\
\Delta\bigg(\usegraph{13}{3305L}\!\bigg)
&\xrightarrow{p_3\to 0}\Delta\bigg(\usegraph{10}{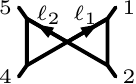}\!\bigg),
\end{align}
\end{subequations}
which can be checked by explicit evaluation.

\subsection{The six-gluon integrand}

With six gluons scattering there are twelve nonzero topologies.
The six genuine two-loop topologies are
\begin{subequations}\label{eq:2L6ga}
\begin{align}
\Delta\bigg(\!\usegraph{10}{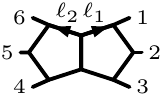}\!\bigg)
&=s_{123}\trp(1(\ell_1-p_1)(\ell_1-p_{12})34(\ell_2-p_{56})(\ell_2-p_6)6)F_1,
\label{eq:441}\\
\Delta\bigg(\usegraph{10}{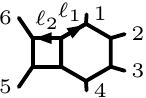}\!\bigg)
&=-s_{56}\trp(123456)\mu_{11}F_1,\label{eq:531}\\
\Delta\bigg(\usegraph{10}{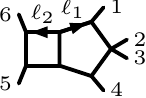}\!\bigg)
&=s_{56}\trp(1(\ell_1-p_1)(\ell_1-p_{123})456)F_1,\label{eq:431M2}\\
\Delta\bigg(\usegraph{10}{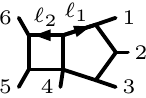}\!\bigg)
&=s_{56}\trp(1(\ell_1-p_1)(\ell_1-p_{12})3(4\!+\!5)6)F_1,\label{eq:4316L}\\
\Delta\bigg(\usegraph{10}{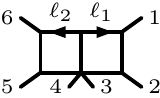}\bigg)&=-s_{12}s_{56}s_{61}F_1,\label{eq:3315LM}\\
\Delta\bigg(\usegraph{13}{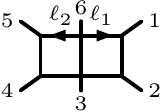}\bigg)&=-s_{12}s_{45}s_{234}F_1.\label{eq:3316L}
\end{align}
\end{subequations}
With the exception of the hexagon box (\ref{eq:531}) these agree with their supersymmetric counterparts.
The hexagon box is a manifestly $D$-dimensional contribution;
to obtain its value we again referred to the $D$-dimensional presentation
of the (parity-even) part of the supersymmetric amplitude given in ref.~\cite{Bern:2008ap}.
The six-leg pentabox (\ref{eq:4316L}) has a counterpart related by symmetry through the horizontal axis,
an expression for which can determined by relabelling the one given above.

The one-loop-squared numerators are
\begingroup
\allowdisplaybreaks
\begin{subequations}\label{eq:2L6gb}
\begin{align}
\Delta\bigg(\!\usegraph{9}{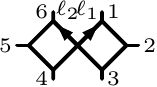}\!\bigg)
&=\trp(1(\ell_1-p_1)(\ell_1-p_{12})34(\ell_2-p_{56})(\ell_2-p_6)6)\times\nonumber\\
&\qquad
\left(F_2+F_3\frac{s_{123}+(\ell_1+\ell_2)^2}{s_{123}}\right),
\label{eq:440}\\
\Delta\bigg(\usegraph{10}{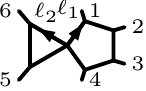}\!\bigg)
&=-\trp(123456)\mu_{11}\left(F_2+F_3\frac{s_{56}+(\ell_1+\ell_2)^2}{s_{56}}\right),
\label{eq:530}\\
\Delta\bigg(\usegraph{10}{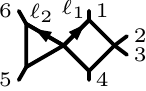}\!\bigg)
&=\trp(1(\ell_1-p_1)(\ell_1-p_{123})456)
\left(F_2+F_3\frac{s_{56}+(\ell_1+\ell_2)^2}{s_{56}}\right).
\label{eq:430M2}\\ 
\Delta\bigg(\usegraph{10}{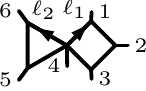}\!\bigg)
&=\frac{\trp(1(\ell_1-p_1)(\ell_1-p_{12})3)}{s_{13}}
\bigg(\text{tr}_+(1356)(F_2+F_3) +\frac{F_3}{s_{123}s_{56}}\times \nn&\qquad\bigg(\trp(134\ell_1\ell_2456)+\big(s_{123}\trp(1356)-s_{56}\trp(1346)\big)(\ell_1+\ell_2)^2 \bigg)\bigg),
\label{eq:4306L}\\
\Delta\bigg(\usegraph{10}{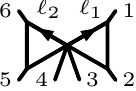}\!\bigg)
&=\trp(1256)(F_2+F_3)+\frac{F_3}{s_{12}s_{123}s_{56}}\bigg(
-s_{12}\trp(134\ell_1\ell_2456)\nonumber\\
&\qquad
-s_{56}\trp(123\ell_1\ell_2346)+s_{123}\trp(12(3\!+\!4)\ell_1\ell_2(3\!+\!4)56)\nonumber\\
&\qquad
+\left(s_{34}+s_{123}\frac{\trp(2345)}{s_{23}s_{45}}\right)\trp(123\ell_1\ell_2456)
+s_{34}\mu_{12}\trp(123456)\nonumber\\
&\qquad
+\Big(s_{12}s_{56}\trp(1346)-s_{123}s_{34}\trp(1256)\nonumber\\
&\qquad\qquad
+s_{123}(s_{12}s_{56}s_{16}+(s_{12}+s_{56})\trp(1256))\Big)(\ell_1+\ell_2)^2\bigg),
\label{eq:3305LM}\\
\Delta\bigg(\usegraph{13}{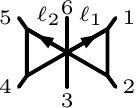}\!\bigg)
&={}\trp(1245)(F_2+F_3)+\frac{F_3}{s_{12}s_{45}s_{123}s_{345}}
\bigg( f_0(123456;\ell_1-p_1,\ell_2-p_5)\nn
&\qquad + f_1(123456; \ell_1-p_1)  - f_1(216543; \ell_1-p_1)\nn
&\qquad - f_1(456123; \ell_2-p_5)  + f_1(543216; \ell_2-p_5)\nn
&\qquad + f_2(123456; \ell_1-p_1,\ell_2-p_5)  +  f_2(456123; \ell_2-p_5, \ell_1-p_1) \nn
&\qquad + f_3(123456; \ell_1-p_1,\ell_2-p_5)\, s_{123}  +  f_3(456123; \ell_1-p_1,  \ell_2-p_5)\, s_{345}\nn
& \qquad + (\ell_1-p_1)^2\, f_4(123456) + (\ell_2-p_5)^2\, f_4(456123)
 \bigg).
\label{eq:3306L}
\end{align}
\end{subequations}
\endgroup
The $s_{13}$ pole in the six-leg box triangle (\ref{eq:4306L}) can be removed using trace identities -
we write the expression this way for compactness.
The graph has a counterpart related by symmetry through the horizontal axis.
The expression for the six-leg double triangle (\ref{eq:3306L}) is somewhat less compact:
the functions $f_i$ can be expressed as
\begin{align}
 f_0(123456;\ell_1,\ell_2) ={}&-s_{123} s_{345} \trp(1245)^2
+ 2 (\ell_1\cdot \ell_2) \Big(s_{12} s_{123} s_{345} \trp(1245)\nn&+s_{45} s_{123} s_{345} \trp(1245)+s_{12} s_{45} s_{345} \trp(1346)\nn&-s_{345} \trp(1236) \trp(4563)-s_{123} \trp(1236) \trp(4563)\nn&+s_{12} s_{45} s_{123} \trp(5326)+s_{12} s_{45} s_{123} s_{234} s_{345}\Big)
\nonumber \\
&-\mu_{12} (s_{123}+s_{345}) \trp(1236) \trp(4563),\\
f_1(123456,\ell) = {}&-s_{123} (s_{12} s_{45} s_{56} \trp(\ell\, 234)+s_{12} s_{34} \trp(\ell\, 23654)\nn&+s_{12} s_{345} \trp(\ell\, 24654)+s_{345} \trp(\ell\, 2451245)),
\\
f_2(123456;\ell_1,\ell_2)={}&s_{123} s_{345} \trp(123\, \ell_1\, \ell_2\, 345)-s_{45} s_{345} \trp(123\, \ell_1\, \ell_2\, 346)\nn&-s_{12} s_{123} \trp(623\, \ell_1\, \ell_2\, 345),
\\
f_3(123456;\ell_1,\ell_2)={}&\trp(123\, \ell_1\, \ell_2\, 65436),
\\
f_4(123456)={}&s_{123} (s_{45} s_{345} \trp(1245)-s_{12} s_{45} \trp(2653)+s_{12} s_{345} \trp(4563)\nn&+s_{12} s_{45} \trp(5326)).
\end{align}
This integrand obeys the soft limits on $p_3$ and $p_6$:
\begin{subequations}
\begin{align}
  &\Delta\bigg(\usegraph{13}{3306L}\!\bigg)\xrightarrow{p_3\to 0}
  \Delta\bigg(\usegraph{9}{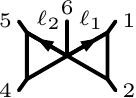}\!\bigg), \\&
  \Delta\bigg(\usegraph{13}{3306L}\!\bigg)\xrightarrow{p_6\to 0}
  \Delta\bigg(\usegraph{13}{3305L}\!\bigg),
\end{align}
\end{subequations}
which can be checked using spinor algebra.

\section{Infrared pole structure}\label{sec:ir}

Since all-plus amplitudes are zero at tree level,
the universal IR structure should be that of a one-loop amplitude \cite{Catani:2000ef}
\begin{align}\label{eq:ircatani}
&A^{(2)}(1^+,2^+,\cdots,n^+)\nonumber\\
&\qquad
=-\sum_{i=1}^n\frac{c_\Gamma}{\epsilon^2}\left(\frac{1}{-s_{i,i+1}}\right)^\epsilon
A^{(1)}(1^+,2^+,\cdots,n^+)+F^{(2)}(1^+,2^+,\cdots,n^+),
\end{align}
where $F^{(2)}$ is finite in the limit $\eps\to0$ and we have reintroduced the standard loop prefactor
\begin{align}\label{eq:cGamma}
c_\Gamma=\frac{r_\Gamma}{(4\pi)^{2-\epsilon}}
=\frac{\Gamma(1+\epsilon)\Gamma^2(1-\epsilon)}{(4\pi)^{2-\epsilon}\Gamma(1-2\epsilon)}.
\end{align}
Reproducing this behaviour requires us to find the IR
divergences up to $\cO(\epsilon^{-1})$ in all of our two-loop integrals.
This we do by following the same approach as in ref.~\cite{Badger:2015lda}.
The two-loop integrals are broken into sums of regions with
soft singularities and evaluated in their respective limits.
The new feature of using local integrands is that the same approach now correctly
predicts the $1/\epsilon$ poles in all cases as well as the leading $1/\epsilon^2$.
It becomes unnecessary to evaluate the resulting one-loop integrals
so eq.~(\ref{eq:ircatani}) can be verified at the integrand level.

\subsection{Soft limits of two-loop integrals \label{sec:softlimits}}

All IR divergent integrals in our two-loop amplitudes come from topologies containing
\begin{align}\label{eq:F1}
F_1=(D_s-2)(\mu_{11}\mu_{22}+(\mu_{11}+\mu_{22})^2
+2\mu_{12}(\mu_{11}+\mu_{22}))+16(\mu_{12}^2-\mu_{11}\mu_{22}).
\end{align}
As the external momenta $p_i$ live in four dimensions,
going into any soft region requires taking the $(-2\epsilon)$-dimensional part of one
of the loop momenta $\ell^{[-2\epsilon]}_i\to0$.
In this limit
\begin{align}
F_1\xrightarrow{\ell^{[-2\epsilon]}_1\to0}(D_s-2)\mu_{22}^2, &&
F_1\xrightarrow{\ell^{[-2\epsilon]}_2\to0}(D_s-2)\mu_{11}^2.
\end{align}
Collinear limits also require $\ell^{[-2\epsilon]}_i\to0$.
The $F_1$ numerator therefore prevents any divergences beyond $\cO(\epsilon^{-2})$
as only one of the loop momenta can enter a soft or collinear region at a time without $F_1$ vanishing.
Further details can be found in ref.~\cite{Badger:2015lda} where this technique has been applied
to the soft singularity of the non-planar five-gluon all-plus amplitude.

Soft singularities always occur between adjacent massless legs in a loop integral.
However, the introduction of local integrands renders many seemingly soft regions finite.
Taking the regulated pentabox integral as an example,
we notice that the integral has only one soft region: $\ell_2\to p_5$.
In this limit,
\begin{align}\label{eq:431ir}
I^D\bigg(\usegraph{10}{431s}\bigg)[F_1]
\xrightarrow{\ell_2\to p_5}(D_s-2)I^D\bigg(\usegraph{9}{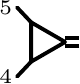}\!\bigg)
I^D\bigg(\!\usegraph{12}{pents}\!\bigg)[\mu^4].
\end{align}
The other two supposedly soft limits $\ell_1\to p_1$ and $\ell_1\to p_{12}$
are finite as the numerator $\trp(1(\ell_1-p_1)(\ell_1-p_{12})3)$ vanishes in these regions -
this is the same phenomenon as we saw in the regulated box integral (\ref{eq:localboxintegral}).

Having numerically evaluated the regulated pentabox integral
using FIESTA's numerical sector decomposition algorithm \cite{Smirnov:2013eza,Smirnov:2015mct},
we find that the decomposition (\ref{eq:431ir}) correctly predicts the IR structure at
$\cO(\epsilon^{-1})$ as well as the leading $1/\epsilon^2$ pole.
The decomposition must therefore be accounting for collinear as well as soft singularities.
This property does not hold true for a scalar pentabox integral
with its additional soft regions $\ell_1\to p_1$ and $\ell_1\to p_{12}$ -
in that case the same technique correctly predicts only the $1/\epsilon^2$ pole.

The only other IR-divergent five-point integral is the five-leg double box.
This integral has two soft regions: $\ell_1\to p_1$ and $\ell_2\to p_5$.
The same procedure reveals that
\begin{align}\label{eq:3315Lir}
&I^D\bigg(\usegraph{13}{3315L}\bigg)[F_1]=-(D_s-2)\times\nonumber\\
&\qquad\left(
I^D\bigg(\usegraph{9}{tri45}\!\bigg)I^D\bigg(\usegraph{10}{box34}\bigg)[\mu^4]+
I^D\bigg(\usegraph{10}{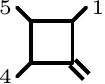}\bigg)[\mu^4]I^D\bigg(\!\usegraph{9}{tri12}\bigg)
\right)+\cO(\epsilon^0).
\end{align}

At six points, the genuine two-loop integrals are
\begin{subequations}\label{eq:6ptir}
\begin{align}
I^D\bigg(\!\usegraph{10}{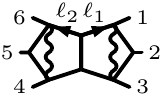}\!\bigg)[F_1]
&=\cO(\epsilon^0),\label{eq:441ir}\\
I^D\bigg(\usegraph{10}{531}\!\bigg)[\mu_{11}F_1]
&=-(D_s-2)I^D\bigg(\usegraph{9}{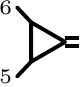}\!\bigg)I^D\bigg(\usegraph{12}{hex}\bigg)[\mu^6]
+\cO(\epsilon^0),\label{eq:531ir}\\
I^D\bigg(\usegraph{10}{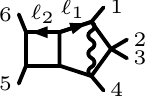}\!\bigg)[F_1]
&=(D_s-2)I^D\bigg(\usegraph{9}{tri56}\!\bigg)
I^D\bigg(\usegraph{12}{pent23s}\!\bigg)[\mu^4]+\cO(\epsilon^0),\label{eq:431M2ir}\\
I^D\bigg(\usegraph{10}{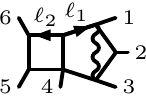}\!\bigg)[F_1]
&=(D_s-2)I^D\bigg(\usegraph{9}{tri56}\!\bigg)
I^D\bigg(\usegraph{12}{pent45s}\!\bigg)[\mu^4]+\cO(\epsilon^0),\label{eq:4316Lir}\\
I^D\bigg(\usegraph{9}{3315LM}\bigg)[F_1]
&=-(D_s-2)\times\nonumber\\
&\!\!\!\!\!\!\!\!\!\!\!\!\!\!\!\!\!\!\!\!\!\!\!\!\!\!\!\!\!\!\!\!\!\!\!\!\!\!\!\!\!\!\left(
I^D\bigg(\usegraph{9}{tri56}\!\bigg)I^D\bigg(\usegraph{10}{box345}\bigg)[\mu^4]
+I^D\bigg(\usegraph{10}{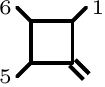}\bigg)[\mu^4]I^D\bigg(\!\usegraph{9}{tri12}\bigg)
\right)+\cO(\epsilon^0),\label{eq:3315LMir}\\
I^D\bigg(\usegraph{13}{3316L}\bigg)[F_1]
&=-(D_s-2)\times\nonumber\\
&\!\!\!\!\!\!\!\!\!\!\!\!\!\!\!\!\!\!\!\!\!\!\!\!\!\!\!\!\!\!\!\!\!\!\!\!\!\!\!\!\!\!\left(
I^D\bigg(\usegraph{9}{tri45}\!\bigg)I^D\bigg(\,\usegraph{10}{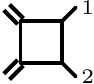}\bigg)[\mu^4]
+I^D\bigg(\usegraph{10}{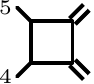}\,\bigg)[\mu^4]I^D\bigg(\!\usegraph{9}{tri12}\bigg)
\right)+\cO(\epsilon^0).\label{eq:3316Lir}
\end{align}
\end{subequations}
where we have checked all topologies with 8 or fewer propagators using the
sector decomposition algorithms implemented in FIESTA~\cite{Smirnov:2013eza,Smirnov:2015mct} and SecDec3.0
\cite{Borowka:2015mxa}. In the ``hexabox'' integral (\ref{eq:531ir}) the extra $\mu_{11}$ term regulates the $\ell_1$ loop.
In this sense it plays a role analogous to the $\trp$ ``wavy line'' structures used in the pentagon integrals.

\subsection{Integrand-level infrared structure}

The IR divergences in the two-loop integrals given above all arise from unregulated box subintegrals.
Therefore, as the one-loop $\mu^4$ integrals are all finite,
the $\epsilon$ poles all come from triangle integrals.
An explicit expression for the triangle integral was given in eq.~(\ref{eq:triintegral}):
its analytic structure is highly reminiscent of the IR operator given in eq.~(\ref{eq:ircatani}).
This observation motivates a simple approach to verifying the universal IR structure.

In the five-point example we proceed as follows.
First we we take the full two-loop amplitude and substitute the IR divergences
given in eqs.~(\ref{eq:431ir}) and (\ref{eq:3315Lir}), summing over cyclic permutations:
\begin{align}
&i\braket{12}\braket{23}\braket{34}\braket{45}\braket{51}(4\pi)^{4-2\epsilon}
A^{(2)}(1^+,2^+,3^+,4^+,5^+)\nonumber\\
&=\sum_{\sigma\in Z_5}\sigma\circ\Bigg(
\frac{s_{45}\trp(1345)}{s_{13}}I^D\bigg(\usegraph{10}{431s}\bigg)[F_1]
-s_{12}s_{45}s_{15}I^D\bigg(\usegraph{13}{3315L}\bigg)[F_1]+\cdots\Bigg)\nonumber\\
&=-(D_s-2)\frac{r_\Gamma}{\epsilon^2}\sum_{\sigma\in Z_5}\sigma\circ\Bigg(
(-s_{45})^{-\epsilon}\frac{\trp(1345)}{s_{13}}
I^D\bigg(\!\usegraph{12}{pents}\!\bigg)[\mu^4]\\
&\qquad
+(-s_{12})^{-\epsilon}s_{45}s_{15}I^D\bigg(\usegraph{10}{box23}\bigg)[\mu^4]
+(-s_{45})^{-\epsilon}s_{12}s_{15}I^D\bigg(\usegraph{10}{box34}\bigg)[\mu^4]
\Bigg)+\cO(\epsilon^0),\nonumber
\end{align}
where we have used the triangle integral (\ref{eq:triintegral}).
Next, we exploit the sum on cyclic permutations to relabel the box integral with a massive $p_{23}$ leg:
\begin{align}
&i\braket{12}\braket{23}\braket{34}\braket{45}\braket{51}(4\pi)^{4-2\epsilon}
A^{(2)}(1^+,2^+,3^+,4^+,5^+)\nonumber\\
&=-(D_s-2)\frac{r_\Gamma}{\epsilon^2}\sum_{\sigma\in Z_5}\sigma\circ(-s_{45})^{-\epsilon}\Bigg(
\frac{\trp(1345)}{s_{13}}I^D\bigg(\!\usegraph{12}{pents}\!\bigg)[\mu^4]
+s_{23}s_{34}I^D\bigg(\usegraph{10}{box51}\bigg)[\mu^4]\nonumber\\
&\qquad
+s_{12}s_{15}I^D\bigg(\usegraph{10}{box34}\bigg)[\mu^4]
\Bigg)+\cO(\epsilon^0),
\end{align}
where we have extracted an overall factor of $(-s_{45})^{-\epsilon}$ inside the sum.
The term in brackets we recognise from eq.~(\ref{eq:5pt1lallplus})
as the planar five-gluon, one-loop all-plus amplitude (multiplied by some extra factors) -
this is of course invariant under cyclic permutations.
Rearranging, we arrive at the desired result:
\begin{align}
A^{(2)}(1^+,2^+,3^+,4^+,5^+)
=-\sum_{i=1}^5\frac{c_\Gamma}{\epsilon^2}\left(\frac{1}{-s_{i,i+1}}\right)^\epsilon
A^{(1)}(1^+,2^+,3^+,4^+,5^+)+\cO(\eps^0),
\end{align}
which agrees with our expectation from eq.~(\ref{eq:ircatani}).

The six-gluon calculation is completely analogous.
By applying the IR singularities given in eqs.~(\ref{eq:6ptir})
to the integrated versions of the six-point numerators presented in eqs.~(\ref{eq:2L6ga})
one can reproduce the one-loop amplitude as presented in eq.~(\ref{eq:6pt1lallplus}).

\section{Rational terms}\label{sec:rational}

The integrated forms of the finite contributions to the leading-colour amplitudes are
\begin{align}\label{eq:A6}
F^{(2)}(1^+,2^+,\cdots,n^+)=(D_s-2)P_n^{(2)}+(D_s-2)^2R_n^{(2)}+\cO(\eps),
\end{align}
where $F^{(2)}$ was introduced in eq.~(\ref{eq:ircatani}).
The rational terms $R_n^{(2)}$ all come from the one-loop-squared topologies
while the polylogarithmic terms $P_n^{(2)}$ come from the topologies shared with $\cN=4$.
These have been identified in a recent computation \cite{Dunbar:2016cxp}.\footnote{
	We have not explicitly checked that the finite part our expressions match those obtained
	in ref.~\cite{Dunbar:2016cxp} owing to the complexity of the multi-scale two-loop integrals appearing.
	The $(\cN=4) \times F_1$ property has been explicitly checked at five points
	and we have no reason to expect different behaviour at six points.}

The rational terms can be found using the one-loop squared integrals listed in appendix \ref{sec:integrals}.
Inserting these expressions into our $D$-dimensional integrands we find that $R_5^{(2)}$
precisely matches the rational part of the full amplitude given in ref.~\cite{Gehrmann:2015bfy}.
At six points we find that $R_6^{(2)}$ can be written as\footnote{
	The integrands of eqs.~(\ref{eq:440}-\ref{eq:3306L}) should be
	combined together with the appropriate symmetry factors.
	The complete expression is available in an ancillary file included in the \texttt{arXiv} submission.}
\begin{align}
R_6^{(2)}=-\frac{1}{144}\frac{i(D_s-2)^2}{\braket{12}\braket{23}\braket{34}\braket{45}\braket{56}\braket{61}(4\pi)^{4-2\epsilon}}
\sum_{\sigma\in Z_6}\sigma\circ\Big(f_R(123456)+f_R(654321)\Big),
\end{align}
which makes its cyclic and reversal symmetry manifest.  The function
$f_R$ can written as a sum of contributions corresponding to
physical pole structures in the external invariants
\begin{align}
f_R(123456) ={} &\frac{2 s_{23} s_{34} s_{45} \trp(1256)}{s_{12} s_{56} s_{123}} + \frac{\trp(1236)}{s_{12} s_{123}}\bigg(-4 s_{23} s_{34}+2 \trp(1345)+6 \trp(2345)\bigg)
\nonumber \\
&+ \frac{1}{s_{123}}\bigg(-12 s_{34} \trp(1236)-s_{34} \trp(1256)\nn
& \qquad +3 (s_{12}+s_{34}+s_{56}) \trp(1346)-16 s_{12} s_{16} s_{34}\bigg)
\nonumber \\
&-\frac{\trp(1245)^2}{s_{12} s_{45}} + \frac{\trp(1256)}{s_{12} s_{56}}\bigg(-2 \trp(p_{14}345)-2 \trp(1256)\bigg)
\nonumber \\
&+\frac{1}{s_{12}}\bigg(2 (s_{16}-s_{34}+s_{45}) \trp(1234)+2 (s_{23}-s_{34}+s_{45}-s_{123}) \trp(1235)\nn
& \qquad +2 (s_{23}+s_{45}-3 s_{123}) \trp(1245)\bigg)
\nonumber \\
&-\frac{1}{4} s_{12} (59 s_{23}-8 s_{34}-56 s_{45})+\frac{1}{4} s_{123} (-4 s_{12}-4 s_{23}+39 s_{34}-40 s_{234}) \nn
& +\frac{9}{4} \trp(1234)+\frac{35}{4} \trp(1235)+\frac{15}{4} \trp(1245).
\end{align}
We have checked that this expression satisfies all universal collinear limits
and agrees with the computation of Dunbar and Perkins \cite{Dunbar:2016gjb}.

\section{Conclusions}\label{sec:conclusions}

In this paper we have explored the use of $D$-dimensional local integrands as a means
of obtaining compact analytic representations of multi-leg two-loop amplitudes. The integrands,
introduced in studies of planar $\mathcal{N}=4$ supersymmetric Yang-Mills, were shown to be powerful
tools in the context of dimensionally regulated amplitudes. As the simplest two-loop amplitudes
in a non-supersymmetric theory we chose the all-plus helicity sector as a testing ground and
found $D$-dimensional local representations of the five- and six-gluon amplitudes.

The representations benefit from highlighting certain physical properties.
The infrared structure is manifest at the integrand level
and the integral coefficients are free of spurious poles.
This results in a reduction in the number of basis integrals,
thereby constraining the overall analytic form of the amplitude.
The rational terms result from one-loop squared topologies and were obtained in a compact form -
at six points this is in agreement with the expressions obtained by Dunbar and Perkins
using augmented BCFW recursion relations~\cite{Dunbar:2016gjb}.

Finding expressions for the one-loop squared topologies, which depend on a six-gluon tree amplitude, proved the most
complicated part of the computation. Though it was possible to find a number of different local representations,
none were as compact as the other topologies. The extremely simple form of the integrated expression
suggests that the integrand expression could yet be improved further. An important additional
check on the integrated expression came from the known collinear limits. It may be that collinear
limits at the integrand level give additional information but would require the development of
additional technology.

General two-loop amplitudes are of course far more complicated than the all-plus amplitudes
considered here. Nevertheless, the techniques we have used should be applicable to the general case
as well. There remain many open questions however: for instance,
one would need to identify a complete basis of local integrands outside of the specific examples considered.

Finally, we note that the present study was restricted to planar amplitudes. Since the nonplanar
sector can be connnected to the planar sector using colour-kinematics relations, 
as was demonstrated at five points in ref. \cite{Badger:2015lda},
the results presented here may be of use in identifying local representations of nonplanar amplitudes.
It would be interesting to see if all-plus amplitudes continue to connect with recent studies in
$\cN=4$ supersymmetric Yang-Mills theory \cite{Arkani-Hamed:2014via,Arkani-Hamed:2014bca,Bern:2014kca,Bern:2015ple}.

\begin{acknowledgments}
  We would like to thank David Dunbar, Johannes Henn, Alexander Ochirov, Donal O'Connell and Warren
  Perkins for useful discussions. We are especially grateful to Christian Br\o{}nnum-Hansen for providing Mathematica code for the
  evaluation of 6D tree amplitudes via BCFW recursion. S.B. is supported by an STFC Rutherford Fellowship ST/L004925/1
  and T.P. is supported by Rutherford Grant ST/M004104/1.  G.M. is supported by an STFC Studentship ST/K501980/1.
\end{acknowledgments}

\appendix
\section{Connection to local integrands in $\mathcal{N}=4$}\label{sec:N4localintegrands}

In this appendix we show how momentum-twistor-based expressions for
planar $\mathcal{N}=4$ local integrands up to two loops,
given in ref.~\cite{ArkaniHamed:2010kv}, may be converted to a form applicable in $D\neq4$.
For instance,
\begin{subequations}\label{eq:arkaniexamples}
\begin{align}
\usegraph{13}{pent23s}
&=\int\displaylimits_{AB}\frac{\braket{1456}\braket{AB|(612)\cap(345)}}
{\braket{AB12}\braket{AB34}\braket{AB45}\braket{AB56}\braket{AB61}},\\
\usegraph{10}{441s}
&=\!\!\!\!\int\displaylimits_{(AB,CD)}\!\!\!\!
\frac{\braket{1346}\braket{AB|(612)\cap(234)}\braket{CD|(345)\cap(561)}}
{\braket{AB61}\braket{AB12}\braket{AB23}\braket{AB34}\braket{CD34}\braket{CD45}\braket{CD56}\braket{CD61}\braket{ABCD}}.
\end{align}
\end{subequations}

Momentum twistors are defined with respect to dual-space coordinates $x_i$,
themselves introduced using $p_i=x_i-x_{i-1}$.
A dual-space point $x$ is identified with a line of twistors, $Z=(\lambda,\mu)$,
satisfying the incidence relation $\mu_{\dot{a}}=\lambda^a x_{a\dot{a}}$.
These twistors form a projective line in $\mathbb{C}\mathbb{P}^3$.
To identify a point $x$ in dual space
it therefore suffices to specify two momentum twistors, $Z_i$ and $Z_j$:
\begin{align}\label{eq:dualspace}
x_{a\dot{a}}=\frac{\lambda_{i,a}\mu_{j,\dot{a}}-\lambda_{j,a}\mu_{i,\dot{a}}}{\braket{ij}}
=x_{i,a\dot{a}}+\frac{\lambda_{i,a}\lambda_j^b(p_{i+1,\ldots,j})_{b\dot{a}}}{\braket{ij}},
\end{align}
where $\braket{ij}\equiv\epsilon_{ab}\lambda_i^a\lambda_j^b$.
The latter identity, using the massless Weyl equation $\lambda_i^ap_{i,a\dot{a}}=0$
implies that when $j=i+1$, $x=x_i$.

As for the loop momenta, for $n$-point scattering at one loop we introduce
$y=\ell+x_n$ associated with the line in $(Z_A,Z_B)$.
At two loops we introduce $y_1=\ell_1+x_n$ and $y_2=-\ell_2+x_n$
associated with the lines $(Z_A,Z_B)$ and $(Z_C,Z_D)$ respectively.\footnote{
  For a complete introduction see ref.~\cite{ArkaniHamed:2010gh}.}

The basic building block to evaluate is the twistor four-bracket,
\begin{align}\label{eq:twistor4bracket}
\braket{i,j,k,l}\equiv\epsilon_{IJKL}Z_i^IZ_j^JZ_k^KZ_l^L
=\braket{ij}\braket{kl}(x-y)^2,
\end{align}
where $x$ and $y$ are associated with the lines $(Z_i,Z_j)$ and $(Z_k,Z_l)$ respectively.
Using the dual-space definition (\ref{eq:dualspace}) it follows that
\begin{subequations}\label{eq:4bracketexamples}
\begin{align}
\braket{ijkl}&=\braket{ij}\braket{kl}
\left(p_{i+1,\ldots,k}^\mu-\frac{\braket{i|\gamma^\mu p_{i+1,\ldots,j}|j}}{2\braket{ij}}
+\frac{\braket{k|\gamma^\mu p_{k+1,\ldots,l}|l}}{2\braket{kl}}\right)^2,\label{eq:4bracketgeneral}\\
\braket{ABij}&=\braket{AB}\braket{ij}
\left(\ell_1^\mu-p_{1,\ldots,i}^\mu-\frac{\braket{i|\gamma^\mu p_{i+1,\ldots,j}|j}}{2\braket{ij}}\right)^2,
\label{eq:4bracketl1}\\
\braket{CDij}&=\braket{CD}\braket{ij}
\left(\ell_2^\mu+p_{1,\ldots,i}^\mu+\frac{\braket{i|\gamma^\mu p_{i+1,\ldots,j}|j}}{2\braket{ij}}\right)^2,
\label{eq:4bracketl2}\\
\braket{ABCD}&=\braket{AB}\braket{CD}(\ell_1+\ell_2)^2.\label{eq:4bracketl1l2}
\end{align}
\end{subequations}
where $i<j<k<l$ (this can always be ensured using antisymmetry of the 4-bracket).
Some frequently-encountered examples are
\begin{subequations}\label{eq:4bracketeasycases}
\begin{align}
\braket{i,i+1,j,j+1}
&=\braket{i,i+1}\braket{j,j+1}s_{i+1,\ldots,j},\label{eq:4bracket1}\\
\braket{A,B,i,i+1}
&=\braket{AB}\braket{i,i+1}(\ell_1-p_{1,\ldots,i})^2,\label{eq:4bracket2}\\
\braket{C,D,i,i+1}
&=\braket{CD}\braket{i,i+1}(\ell_2+p_{1,\ldots,i})^2,\label{eq:4bracket3}
\end{align}
\end{subequations}
where the massless Weyl equation is again used to make these simplifications.

When evaluating four-brackets involving intersections of planes in momentum-twistor space
one should use the twistor intersection formula,
\begin{align}\label{eq:twistorintersection}
\braket{AB|(abc)\cap(def)}
=\braket{cdef}\braket{ABab}+\braket{adef}\braket{ABbc}+\braket{bdef}\braket{ABca}.
\end{align}
From this we have established the general pattern that
\begin{align}
\usegraph{13}{pentschemes}
&\sim\braket{AB|(i-1,i,i+1)\cap(j-1,j,j+1)}\nonumber\\
&\qquad
=-\braket{AB}\braket{i-1,i}\braket{i,i+1}\braket{j-1,j}\braket{j,j+1}\left[i|\ell_x\ell_y|j\right],
\end{align}
which is manifestly local.
Application of $\trp(i\ell_x\ell_yj)=\left[i|\ell_x\ell_y|j\right]\braket{ji}$ gives
agreement with the integral definition we made in eq.~(\ref{eq:squiggledef}).
For instance, in the two examples given in eqs.~(\ref{eq:arkaniexamples}),
\begin{subequations}
\begin{align}
\braket{AB|(612)\cap(345)}
&=\frac{\braket{AB}\braket{61}\braket{12}\braket{34}\braket{45}}{\braket{14}}
\trp(1(\ell-p_1)(\ell-p_{123})4),\\
\braket{AB|(612)\cap(234)}
&=\frac{\braket{AB}\braket{61}\braket{12}\braket{23}\braket{34}}{\braket{13}}
\trp(1(\ell_1-p_1)(\ell_1-p_{12})3),\\
\braket{CD|(345)\cap(561)}
&=\frac{\braket{CD}\braket{34}\braket{45}\braket{56}\braket{61}}{\braket{46}}
\trp(4(\ell_2-p_{56})(\ell_2-p_6)6).
\end{align}
\end{subequations}

When calculating complete integrands in $D\neq4$ one should use the measure correspondences
\begin{align}\label{eq:twistormeasure}
\int\displaylimits_{AB}\frac{1}{\braket{AB}^4}\sim\int d^4\ell,&&
\int\displaylimits_{(AB,CD)}\frac{1}{\braket{AB}^4\braket{CD}^4}
\sim\int d^4\ell_1d^4\ell_2,
\end{align}
at one and two loops repectively.
This accounts for all factors of $\braket{AB}$ and $\braket{CD}$.
An overall factor of the tree amplitude $\mathcal{A}^{(0),[\mathcal{N}=4]}_\text{MHV}$ then contributes both a Parke-Taylor denominator
$\braket{12}\cdots\braket{n-1,n}\braket{n1}$ and the supersymmetric delta function $\delta^8(Q)$.

\section{Two-loop integrals \label{sec:integrals}}

This appendix contains all the integrals required to compute the
rational part of the five- and six-point all-plus amplitudes. These
expressions are also available in an ancillary file included with the
\texttt{arXiv} submission.

\subsection{Five-point integrals}

\begin{subequations}
\begin{align}
I^{4-2\epsilon}\bigg(\usegraph{13}{430s}\bigg)[\mu_{11}\mu_{22}]
={}&\frac{s_{13}}{4}+\cO(\epsilon),\\
I^{4-2\epsilon}\bigg(\usegraph{13}{430s}\bigg)[\mu_{11}\mu_{22}(\ell_1+\ell_2)^2]
={}&\frac{\trm(1345)-s_{13}(6s_{45}-2s_{13}-s_{34}-s_{15})}{36}\nn&+\cO(\epsilon),\\
I^{4-2\epsilon}\bigg(\usegraph{13}{3305L}\bigg)[\mu_{11}\mu_{22}]
={}&\frac{1}{4}+\cO(\epsilon),\\
I^{4-2\epsilon}\bigg(\usegraph{13}{3305L}\bigg)[\mu_{11}\mu_{22}\text{tr}_+(123\ell_1\ell_2345)]
={}&\frac{\trp(123(2p_1\!+\!p_2)(p_4\!+\!2p_5)345)}{36}+\cO(\epsilon),\\
I^{4-2\epsilon}\bigg(\usegraph{13}{3305L}\bigg)[\mu_{11}\mu_{22}(\ell_1+\ell_2)^2]
={}&\frac{(2p_1\!+\!p_2)\cdot(p_4\!+\!2p_5)}{18}+\cO(\epsilon).
\end{align}
\end{subequations}

\subsection{Six-point integrals}

\begingroup
\allowdisplaybreaks
\begin{subequations}
\begin{align}
I^{4-2\epsilon}\bigg(\!\usegraph{9}{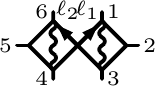}\!\bigg)[\mu_{11}\mu_{22}]
={}&\frac{s_{13}s_{46}}{4}+\cO(\epsilon),\\
I^{4-2\epsilon}\bigg(\!\usegraph{9}{440s}\!\bigg)[\mu_{11}\mu_{22}(\ell_1+\ell_2)^2]
={}&-\frac{s_{13}s_{46}(s_{23}+s_{45})}{12}\nn&
+\frac{s_{13}s_{46}(2p_1\!+\!p_2\!+\!p_3)\cdot(p_4\!+\!p_5\!+\!2p_6)}{18}\nn
&+\frac{s_{46}\trp(12(p_4\!+\!p_5\!+\!2p_6)3)}{36}\nn&+\frac{s_{13}\trp(4(2p_1\!+\!p_2\!+\!p_3)56)}{36}\nn&+\frac{\trp(125643)}{36}+\cO(\epsilon),\\
I^{4-2\epsilon}\bigg(\usegraph{9}{530}\!\bigg)[\mu_{11}^2\mu_{22}]
={}&\cO(\epsilon),\\
I^{4-2\epsilon}\bigg(\usegraph{9}{530}\!\bigg)[\mu_{11}^2\mu_{22}(\ell_1+\ell_2)^2]
={}&-\frac{1}{12}+\cO(\epsilon),\\
I^{4-2\epsilon}\bigg(\usegraph{9}{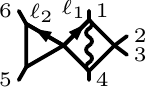}\!\bigg)[\mu_{11}\mu_{22}]
={}&\frac{s_{14}}{4}+\cO(\epsilon),\\
I^{4-2\epsilon}\bigg(\usegraph{9}{430M2s}\!\bigg)[\mu_{11}\mu_{22}(\ell_1+\ell_2)^2]
={}&\frac{\trm(1456)-s_{14}(5s_{23}+6s_{56}-2s_{14}-s_{45}-s_{16})}{36}\nn&+\cO(\epsilon),\\
I^{4-2\epsilon}\bigg(\usegraph{9}{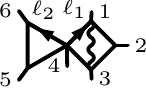}\!\bigg)[\mu_{11}\mu_{22}]
={}&\frac{s_{13}}{4}+\cO(\epsilon),\\
I^{4-2\epsilon}\bigg(\usegraph{9}{4306Ls}\!\bigg)[\mu_{11}\mu_{22}(\ell_1+\ell_2)^2]
={}&-\frac{s_{13}s_{23}}{12}+\frac{\trp(12(p_5\!+\!2p_6)3)}{36}\nn
&+\frac{s_{13}(2p_1\!+\!p_2\!+\!p_3)\cdot(p_5\!+\!2p_6)}{18}+\cO(\epsilon),\\
I^{4-2\epsilon}\bigg(\usegraph{9}{4306Ls}\!\bigg)[\mu_{11}\mu_{22}\ell_1^\mu\ell_2^\nu]
={}&\frac{(2s_{13}(2p_1\!+\!p_2\!+\!p_3)^\mu+\trp(12\gamma^\mu3))(p_5\!+\!2p_6)^\nu}{72}\nn&+\cO(\epsilon),\\
I^{4-2\epsilon}\bigg(\usegraph{9}{3305LM}\!\bigg)[\mu_{11}\mu_{22}]
={}&\frac{1}{4}+\cO(\epsilon),\\
I^{4-2\epsilon}\bigg(\usegraph{9}{3305LM}\!\bigg)[\mu_{11}\mu_{22}(\ell_1+\ell_2)^2]
={}&\frac{(2p_1\!+\!p_2)\cdot(p_5\!+\!2p_6)}{18}+\cO(\epsilon),\\
I^{4-2\epsilon}\bigg(\usegraph{9}{3305LM}\!\bigg)[\mu_{11}\mu_{22}\ell_1^\mu\ell_2^\nu]
={}&\frac{(2p_1\!+\!p_2)^\mu(p_5\!+\!2p_6)^\nu}{36}+\cO(\epsilon),
\end{align}
\begin{align}
I^{4-2\epsilon}\bigg(\usegraph{13}{3306L}\!\bigg)[\mu_{11}\mu_{22}]
={}&\frac{1}{4}+\cO(\epsilon),\\
I^{4-2\epsilon}\bigg(\usegraph{13}{3306L}\!\bigg)[\mu_{11}\mu_{22}(\ell_1+\ell_2-p_1-p_5)^2]
={}&\frac{(p_1\!-\!p_2)\cdot(p_5\!-\!p_4)}{18}\nn&-\frac{s_{12}+s_{45}}{12}+\cO(\epsilon),\\
I^{4-2\epsilon}\bigg(\usegraph{13}{3306L}\!\bigg)[\mu_{11}\mu_{22}(\ell_1-p_1)^\mu]
={}&\frac{p_2^\mu-p_1^\mu}{12}+\cO(\epsilon),\\
I^{4-2\epsilon}\bigg(\usegraph{13}{3306L}\!\bigg)[\mu_{11}\mu_{22}(\ell_2-p_5)^\mu]
={}&\frac{p_4^\mu-p_5^\mu}{12}+\cO(\epsilon),\\
I^{4-2\epsilon}\bigg(\usegraph{13}{3306L}\!\bigg)[\mu_{11}\mu_{22}(\ell_1-p_1)^\mu(\ell_2-p_5)^\nu]
={}&\frac{(p_1^\mu-p_2^\mu)(p_5^\nu-p_4^\nu)}{36}+\cO(\epsilon),\\
I^{4-2\epsilon}\bigg(\usegraph{13}{3306L}\!\bigg)[\mu_{11}\mu_{22}(\ell_1-p_1)^2]
={}&-\frac{s_{12}}{12}+\cO(\epsilon),\\
I^{4-2\epsilon}\bigg(\usegraph{13}{3306L}\!\bigg)[\mu_{11}\mu_{22}(\ell_2-p_5)^2]
={}&-\frac{s_{45}}{12}+\cO(\epsilon).
\end{align}
\end{subequations}
\endgroup

\bibliographystyle{JHEP}
\bibliography{references}

\end{document}